\documentclass[12pt]{iopart}
\usepackage{graphicx}% Include figure files
\usepackage{bm,wasysym,textcomp}% bold math
\begin{document}
	
	\title{Regimes of electronic transport in doped InAs nanowire.}
	
	\author{A A Zhukov$^1$ and I E Batov$^{1,2}$ }
		%		Ch Bl\"omers$^2$, 
		
		%		, R. Calarco$^2$\footnote{Present address:
		%	Istituto per la Microelettronica E Microsistemi (IMM), 
		% Consiglio Nazionale delle Ricerche,  Via del Fosso del %Cavaliere 100, 00133, Rome, Italy.} 
	\address{ $^1$ Institute of Solid State Physics, Russian Academy of
	Science, Chernogolovka, 142432 Russia}
	\address{ $^2$ Faculty of Physics, National Research University Higher School of 
		Economics, Moscow 101000, Russia}
	\ead{azhukov@issp.ac.ru}
	\date{\today}
	
	\begin{abstract}
		We report on the low temperature measurements of the magnetotransport in Si-doped InAs quantum wire in the presence of a charged tip of an atomic force microscope serving as a mobile gate, i.e. scanning gate microscopy (SGM). By altering the carrier concentration with back gate voltage, we transfer the wire through several transport regimes: from residual Coulomb blockade to nonlinear resonance regime, followed by linear resonance regime and, finally, to almost homogeneous diffusion regime. We demonstrate direct relations between patterns measured with scanning gate microscopy and spectra of universal conductance fluctuations. A clear sign of fractal behavior of magnetoconductance dependence is observed for non-linear and linear resonance transport regimes.
	\end{abstract} 
	\pacs{73.23.Hk, 73.40.Gk, 73.63.Nm}
	
	\maketitle
	
	\section{Introduction}
	
	One of the peculiarities of one-dimensional and quasi-one-dimensional diffusion electronic transport is the presence in these type of systems an electron scattering from so-called resonance scatters. These scatters, such as weak links or any other potential barriers, influence on electrons in all channels. If these scatters are strong enough and temperature $T$ is low, the transport can demonstrate Coulomb blockade \cite{Datta} ($h/\tau,k_BT<e^2/c$) or zero-bias anomaly \cite{Altshuler1985book}, here $h$ is Planck's constant, $k_B$ is Boltzmann's constant, $\tau$ is the tunneling time through barriers, $e$ is elementary charge and $c$ is capacitance of the system section in between two nearest weak links. In the case that potential barriers are weak, the transport is linear but well-defined resonances in magnetoresistance can be observed due to over-barrier reflections \cite{Zhukov2022}. Finally, if the role of such barriers/scatters is negligible, the homogeneous diffusive transport regime is realized \cite{ImryBook}. 
	
	In samples with the phase coherence length ($l_\phi$) comparable to the sample size ($l_{wire}$ the universal conductance fluctuations (UCF) are observed \cite{Altshuler1985}. Using the correlator $F(\Delta B)=\left\langle \delta R^{-1}(B)\delta R^{-1}(B+\Delta B)\right\rangle$, where $\delta R^{-1}(B)=\left\langle R^{-1} \right\rangle-R^{-1}(B)$, it is possible to extract the value of $l_\phi=\Phi_0/dB_c$ from the correlation field $B_c$ defined as $F(B_c)=0.5F(0)$ \cite{Lee1987, Beenakker1988}, $\Phi_0=e/h$ is the quantum flux and $d$ is a constant of the order of unity \cite{Beenakker1988}. 
	
	According to the work \cite{Ketzmerick1996}, the deviation of correlation function $\Delta F(\Delta B) = F(0) - F(\Delta B) \propto \Delta B^\gamma$, and the value of exponent $\gamma$ defines the dimension of the fractional Brownian motion $D=2-\gamma/2$ \cite{MandelbrotBook,Jannsen1995, Evers2008} of magnetoconductance dependence. Fractional Brownian motion has been found in quasiballistic Au-nanowires \cite{Hegger1996}, different types of semiconductor nanowires \cite{Marlow2006}, and graphene stripes \cite{Amin2018}.
	
	Here, we present a detailed investigation of magnetotransport in Si-doped InAs nanowire using SGM mapping and standard magnetotransport measurements in a wide range of carrier concentrations. The system passed through four transport regimes from residual Coulomb blockade to non-linear resonance scattering, then to linear resonance scattering and finally to quasi-homogenious diffusion regime. The role of resonant scatters in formation of fractional Brownian motion of magnetoconductance dependence is discussed.
	
	\section{Experimental}
	
	The nanowires were grown on GaAs (111)B substrates by low-pressure metal organic vapor phase epitaxy (MOVPE). Nitrogen gas (N$_2$) was used as the carrier gas to transport trimethylindium (TMIn) and arsine (AsH$_3$) in the reactor at a working pressure of 20 mbar and a total gas flow rate of 3100 ml/min. The growth temperature was $650 ^\circ$C. For silicon doping during growth, the disilane (Si$_2$H$_6$) flux was adjusted to achieve various n-type doping levels. To quantify the supply of doping species more easily, doping factor, consisting of the partial pressure ratio of dopant versus group III precursor, is defined as a ratio of p(Si$_2$H$_6$)/p(TMIn). The ratio of $7.5 \times 10^{-4}$ was used for growth the batch of the nanowires investigated in the current experiment. Additional details of the Si-doped nanowires growth procedure can be found elsewhere \cite{Wirths2011, Akabori2009}.
	
	The diameter of the wire was $d_{wire}=100$\,nm. The wire was placed on an $n$-type doped Si (100) substrate covered by a thermal grown 100~nm thick SiO$_2$ insulating layer. The Si substrate served as the back-gate electrode. The Ti/Au contacts to the wire as well as the markers of the search pattern were defined by electron-beam lithography. The distance between the contacts was $l_{wire}=2.9\mu$m. A scanning electron beam micrograph of the sample is shown in Fig.~1a). The source and drain metallic electrodes connected to the wire are marked as S and D.
	
	All measurements were performed at $T=\,4.2$\,K. The charged tungsten tip of a home-built scanning probe microscope \cite{AFM} was used as a mobile gate during scanning gate imaging measurements, see Fig.~1b). All scanning gate measurements were performed by keeping the potential of the scanning probe microscope tip ($V_{t}$=0\,V) as well as the back-gate voltage ($V_{BG}$) constant. The conductance of the wire during the scan was measured in a two-terminal circuit by using a standard lock-in technique.  The tip to SiO$_2$ surface distance chosen for the scanning process was $h_{tip}=250$\,nm. During SGM scans and linear magnetotransport measurements a driving AC current with an amplitude of $I_{AC}=0.8$\,nA at a frequency of 231\,Hz was applied while the voltage was measured by a differential amplifier. External magnetic field was directed perpendicular to the wire axis and Si substrate surface.
	
	\section{Experimental results}
	
	In Fig.~2 the dependence of conductance ($R^{-1}$) of the nanowire as a function of back gate voltage is presented. The overall linear dependence of conductance on back gate voltage (carrier density) is typical for Si-doped InAs nanowires \cite{Wirths2011}. Non-regular fluctuations are universal conductance fluctuations (UCF) which arise because the phase coherence length in InAs is comparable to the length of the wire. The magnified low back gate section of $R^{-1}(V_{BG})$ curve is shown in the top-left inset of Fig.~2. Additionally to the UCF, there are oscillations with period of $\Delta V_{BG}=36$\,mV, marked with blue oval. These oscillations come from the residual Coulomb blockade of a quantum dot positioned in the mid of the nanowire \cite{Weis2014}.
	
	According to finite element calculations of the capacitance of the nanowire \cite{Wunnicke2006}, the specific capacitance of our sample is $c_s\simeq 80$pF/m. Thus, it is possible to calculate the carrier density of the nanowire as a function of the back gate voltage $n_e=c_s(V_{BG}-V_{Th})/\pi e(d_{wire}/2)^2$, here $V_{Th}=-5.2$\,V is the threshold voltage, and the elastic mean free path $l_e=(h l_{wire}/2\pi^2 e^2 R(d_{wire}/2)^2)(2\pi^2/n_e^2)^{1/3}$. The low-right inset in Fig.~2 shows the dependencies of $l_e$ (black curve, left scale) and $l_ek_F$ (blue curve, right scale) on back-gate voltage, here $k_F=(3\pi^2n_e)^{1/3}$ is the Fermi vector. Two green arrows point to the $l_e$ curve at $V_{BG}=-1$\,V and 1\,V. It is worth noting that the variations of $l_e$ and $l_ek_F$ as $V_{BG}$ changes from -1\,V to 1\,V is just 25-30\%. 
	
	The results of SGM mapping of the nanowire are presented in Figs.~3\,--\,8. Each figure demonstrates the evolution of the scanning results due to weak variation of the back gate voltage with a step of 10\,mV (Figures a) to f)), and applied external magnetic field ($B=0.1$\,T, 0.2\,T, 0.3\,T, 0.5\,T, 1\,T, and 2\,T, Figures g) to l)) for six back gate voltages, i.e. $V_{BG}=-3$\,V, $-2.1$\,V, $-1$\,V, 1\,V, 4\,V, and 10\,V, respectively. The step of $V_{BG}$ of 10\,mV was chosen to exceed correlation back gate voltage $V_{cBG}=hD/el_\phi^2<5$\,mV for all applied $V_{GB}$, here $D$ is the diffusion coefficient.
	
	Mapping results of the SGM scanning in the vicinity of $V_{BG}=-3$\,V presented in Fig.~3 demonstrate quite complex structure typical for nanowires or nanotubes with a number of the blocking barriers of different opacity \cite{Woodside2002}. The key feature of scans in Fig.~3 is well defined concentric ovals, see Figs.~3f) and 3l) for example. The presence and the position of such ovals means the formation of the quantum dot in the mid-section of the nanowire and ovals are the result of the Coulomb blockade realized in this dot \cite{Woodside2002}. The period of Coulomb blockade oscillations of $\Delta V_{BG}=36$\,mV obtained from $R^{-1}(V_{BG})$ curve allows to estimate the length of electronic system of this quantum dot $l_{dot}\simeq e/c_s \Delta V_{BG} \simeq 60$\,nm. Taking into account the lengths of depletion zones of around 100\,nm, the distance between two strong blocking barriers forming this dot can be estimated as $l_{BB}\simeq 250$\,nm. Thus, this quantum dot is positioned approximately in the center of the wire with barrier-to-barrier distance $l_{BB}\ll l_{wire}$. It means that the InAs nanowire is divided into three sections. The Section I extends from the source contact to the first blocking barrier, the Section II is the quantum dot itself and the Section III lays between the second blocking barrier and the drain contact.
	
	The scanning gate microscopy mapping performed in the vicinity of $V_{BG}=-2.1$\,V is shown in Fig.~4. At this back gate voltage the SGM mapping results look still quite complex, see Fig.~4a), but they demonstrate a response from all three sections of the nanowire. Here equicapacitance ovals and circles are not resulted from the Coulomb blockade, because the shape of them changes dramatically with application of external magnetic field, see Figs.~4g)\,--\,4l). The equicapacitance ovals and circles in SGM data come from the alteration of the opacity of the blocking barriers or, probably, due to the variation of the density of states  \cite{Altshuler1985book}.
	
	Fig.~5 shows the scanning gate microscopy mapping carried out in the vicinity of $V_{BG}=-1$\,V. No well-defined response from each of the Sections can be resolved, but the mapping results are extremely stable against the back gate voltage variation and application of magnetic field up to $B=0.3$\,T, see Figs.~5a)\,--\,5h). Higher magnetic fields from $B=0.5$\,T to 2\,T deform the SGM mapping results, similar to Fig.~4, see Figs.~5j)\,--\,5l).
	
	The scanning gate microscopy mapping done in the vicinity of $V_{BG}=1$\,V is presented in Fig.~6. Both the variation of back gate voltage and the magnetic field as small as 0.1\,T alter the SGM mapping pictures. The smallest scale features of SGM scans have the size of 250\,--\,300\,nm. This size is comparable to the tip to SiO$_2$ surface distance and it is the spatial resolution of the current experimental setup \cite{Zhukov2014SWComp}. Three well-defined resistance minima are visible in scans and marked with arrows in Fig.~6c). As it has been estimated previously, the distance between two barriers forming the central quantum dot is 250\,nm and each of the barrier cannot be resolved separately, so they are visualized as a single one. Thus, these three minima are related to the source contact barrier, quantum dot double barrier, and the drain contact barrier. The positions of them are stable against variation of $V_{BG}$ (Figs.~6a)\,--\,6f)). The double barrier sign (the red spot marked with arrow) is visible even at SGM scan done at $B=2$\,T, see Fig.~6l). The scans shown in Fig.~6 additionally confirm the position of the quantum dot allocated from SGM mapping, see Fig.~4.
	
	The SGM mapping done in the vicinity of $V_{BG}=4$\,V and 10\,V is shown in Fig.~7 and 8, respectively. These scans demonstrate the interplay of the features come from UCF which vary their positions and the residual impact of the blocking barriers on the wire conductivity. It is worth noting that some minor influence of these barriers in SGM scans remains both at $V_{BG}=4$\,V (Fig.~7i)) and at $V_{BG}=10$\,V (Fig.~8a)). This behavior of blocking barrier influence on SGM mapping is the same as observed in InN nanowires previously \cite{Zhukov2022}. 
	
	Figure 9 shows the dependence of the wire conductance ($R^{-1}$) on magnetic field $B$ ($-1$\,T$\leq B\leq 7$\,T) measured at  $V_{BG}=-3$\,V, $-2$\,V, $-1$\,V, 1\,V, 4\,V, and 10\,V, i.e almost at the same values of the back gate voltages as for SGM scans. Curves, except one measured at $V_{BG}=$10\,V, are shifted for clarity. Non-regular reproducible oscillations are UCF. The peak in conductivity around $B=0$\,T for $V_{BG}=$10\,V comes from weak antilocalization quantum correction due to spin-orbit interaction in InAs and it transforms to the localization dip for $V_{BG}\leq-2$\,V. This transition from weak anilocalization to localization with decreasing the carrier concentration is typical for InAs nanowires \cite{Bleszynski2005, Dhara2009, Roulleau2010, Bloemers2011, Boyd2011a, Wang2015}.
	
	The top panel of Fig.~10 shows universal conductance fluctuations spectra calculated from $R^{-1}(B)$ data measured at $V_{BG}=-3$\,V, $-2$\,V, and $-1$\,V. The bottom panel of Fig.~10 presents the spectra measured at $V_{BG}=$1\,V, 4\,V, and 10\,V. The essential suppression of the fluctuation spectra for $1/B>1$\,(T$^{-1}$) for top panel data in comparison to bottom panel data is clearly visible.
	
	The normalized correlator dependencies $F(\Delta B)/F(0)$ calculated from magnetoconductance data obtained at $V_{BG}=-3$\,V, $-2$\,V, $-1$\,V, 1\,V, 4\,V, and 10\,V are integrated in Fig.~11. The smallest value of the correlation field $B_c=0.11$\,T extracted from data measured at $V_{BG}=$10\,V gives $l_\phi=200$\,nm. This value of the phase correlation length is within the range from 200\,nm to 500\,nm of previously obtained values \cite{Bleszynski2005, Dhara2009, Roulleau2010, Bloemers2011, Boyd2011a, Wang2015}. The significant increase of $l_\phi$ value at higher carrier density has been obtained previously as well \cite{Bleszynski2005, Dhara2009, Roulleau2010}. The correlator $F(\Delta B)/F(0)$ is calculated from $R^{-1}(V_{BG})$ data measured in magnetic fields from $B=0.65$\,T to 7\,T to eliminate any influence of low-field quantum corrections.
	
	The dependencies of deviation of the correlator $\Delta F(\Delta B)=F(\Delta B)-F(0)$ in double logarithmic scale for $V_{BG}=-3$\,V, $-2$\,V, $-1$\,V, 1\,V, 4\,V, and 10\,V  are shown in Fig.~12. Two straight lines show the slopes of $\Delta B^\gamma$ dependence for two exponent values $\gamma=1.5$ (red) and $\gamma=2.0$ (black).
	
	The normalized differential resistance $(dV/dI)/(dV/dI(0))$ as a function of driving current measured at $V_{BG}=-3$\,V, $-2$\,V, $-1$\,V, and 1\,V are shown in Fig.~13. The dots on the curves mark values of current where $eV_{SD}=k_BT$. The well-defined non-linear behavior is seen for $V_{BG}=-3$\,V and $-2$\,V only. This experiment is done at $B=0.65$\,T to eliminate the quantum corrections influence.
	
	\section{Discussion}
	
	Transitions from Coulomb blockade to Fabri-P\'erot interference in ballistic one-dimensional and quasi-one-dimensional systems were recently observed and discussed in detail in \cite{Makarovski2007, Wang2019}. In inhomogeneous diffusive quasi-one-dimensional systems, the transport phenomena are more complex. Presented data help to illustrate all relevant transport regimes focusing on their peculiarities in magnetotransport data and the SGM mapping results obtained in the same run.
	
	The doped InAs nanowire is chosen because of its homogeneous radial carrier density \cite{Wirths2011}. Thus, any influence on the transport resulted from cylindrical shape of the electronic system \cite{Luth2010} is minor.  
	
	As it has been mentioned previously, the dependence $R^{-1}(V_{BG})$ (Fig.~2) and SGM data (Fig.~3) obtained at $V_{BG}=-3$\,V allow to estimate the position and the size of the quantum dot formed in the mid-section of the wire. The spectrum of the UCF (see Fig.~10, top panel) confirms the statement that the dominating role belongs to the small area loops ($90\times 50$\,nm$^2$). The transport at this value of the back gate voltage is non-linear, see Fig.~13.
	
	By increasing the carrier density concentration ($V_{BG}=-2$\,V), we transfer the system to non-linear resonant regime (see Fig.~13). In this regime, the strongest resonant scatters (blocking barriers) define three quasi-localized states formed in corresponding Sections of wire. Maps obtained with the SGM technique depend weakly on variation of the back gate voltage and application of the small magnetic field, see Fig.~3. New energy scale of the order of 100\,meV is the characteristic energy of variation of opacity of the blocking barriers and fluctuation of the density of states. The contribution from small area loops is dominant in the UCF spectra (Fig.~10, top panel).

	Further increase of the carrier concentration ($V_{BG}=-1$\,V) gives rise to the reduced influence of the blocking barriers. Transport becomes linear, see Fig.~13. No well-defined patterns allocating three Sections of the nanowire can be observed in SGM scans, but the resonant reflections of electrons still play an important role altering the local carrier density of states and ruling the weak dependence of the SGM mapping results on the back gate voltage and the magnetic field (Fig.~4), similarly to previously described non-linear resonant regime. Contribution from small area loops is still dominant in the UCF spectrum (Fig.~10, top panel). The size of the area of these loops allows to estimate the maximum characteristic length of segments into which the nanowire is divided $l_{seg}\sim 50-100$\,nm .
		
	At $V_{BG}=1$\,V we observe important changes in the SGM scans. The SGM mapping results (Fig.~5) become sensitive to the small (10\,mV) increase of the back gate voltage as well as to application of the small ($B=0.1$\,T) external magnetic field. This situation is similar to the one obtained previously in undoped InAs nanowires \cite{Zhukov2014Comp} and can be considered as diffusive regime with diminished influence of resonant scatters. Additionally, the spectrum of the UCF extends to the higher frequencies, thus the larger area loops start to come into play (Fig.~10, bottom panel). Transport is linear as indiated in Fig.~13.
	
	Further increasing the back gate voltage results in even more homogeneous transport according to SGM data (Figs.~6 and 7) with wide spectra of UCF (Fig.~10, bottom panel). At $V_{BG}=10$\,V correlation field $B_c$ has reached the value related to the phase coherence length of 200\,nm, finally.
	
	The transiton from linear resonant regime to more homogenous diffusive one occurs at back gate voltages between $V_{BG}=-1$\,V and 1\,V (see two arrows in the bottom-right inset of Fig.~2). As it has been mentioned previously, no essential variations of $l_e$ or $k_Fl_e$ happen in this range of back gate voltages. Therefore, it is not possible to find any sign of this transition just from $R^{-1}(V_{BG})$ dependence. But the correlated switch of the behavior of SGM mapping results, as well as sudden increase of the high frequency spectrum of UCF can be considered as an undoubtful evidence of this transition. 
	
	The most exciting feature of the obtained data is non-trivial behavior of the exponent $\gamma$ in $\Delta F(\Delta B) \propto \Delta B^\gamma$ dependencies obtained for different back gate voltages, see Fig.~13. This exponent is $\gamma=2.1\pm 0.1$ for Coulomb blockade regime when the transport is defined mostly by quantum dot. Then, the value of exponent drops to values close to $\gamma=1.4$ in non-linear and linear resonant scattering regimes, see Table~1. The transport depends strongly on resonant scatters in the wire forming a set of quasilocalized states resulting in nonhomogeneous fractional Bownian motion of magnetoconductance dependence with dimensionality greater than one ($Dim=2-\gamma/2=1.3$) \cite{Ketzmerick1996, Hegger1996} for both non-linear and linear resonant regimes. The reason of weak variation of $\gamma$ in this regime is presently not clear and might be the subject for further investigations. At $V_{BG} \geq 1$\,V when the role of the resonance scatters diminishes, the value of $\gamma$ starts to rise toward to the standard value of 2, see Table.~1.
	\\
	
	\begin{table}[ht]
		%\centering
		\caption{The dependence of the correlation field $B_c$ and $\gamma$ on back gate voltage.}
		\begin{tabular}[t]{lcc}
			\hline
			$V_{BG}(V)$&$B_c$(T)&$\gamma$\\
			\hline
			-3	& 0.48 & 2.1$\pm 0.1$\\ 
			-2	& 0.25 & 1.38$\pm 0.1$\\ 
			-1	& 0.27 & 1.42$\pm 0.1$\\ 
			1	& 0.21 & 1.78$\pm 0.1$\\ 
			4	& 0.21 & 1.71$\pm 0.1$\\ 
			10	& 0.11 & 1.90$\pm 0.1$\\
			\hline
		\end{tabular}
	\end{table}%
	
	According to the underlying paper by Altshuler, Gefen, Kamenev and Levitov (AGKL), there are three regimes of many-body localization \cite{Altshuler1997, Mirlin1997}. The first one is realized when the energy $k_BT$ is less than the characteristic energy of interacting quantum dot $E_{QD}$  ($E_{QD}(V_{BG}=-3$\,V$) =e\Delta V_{BG}\sqrt{g/ln(g)} \sim 10$\,meV, and it is the Coulomb blockade regime in this paper). In this regime, one-particle states are very similar to the exact many-body states. Here $g=h/e^2R_{QD} \sim 1$ is the normalized conductance of the central quantum dot. In the {\it intermediate regime} when $k_BT<E_{Th,seg} \sim 1.7$\,meV at $V_{BG}=-1$\,V, the quasiparticle states are delocalized, but they are fractal and non-ergodic \cite{Altshuler1997, Mirlin1997}. Here $E_{Th,seg}=hD/l_{seg}^2$ is the Thouless energy of the typical largest segment of the wire formed by resonant scatters. 
	
	Fractal structure of wave function of the electronic system and overall non-ergodic behavior of this regime correlates well with in the current experiment at non-linear and linear resonant regimes demonstrating fractional Bownian motion of magnetoconductance behavior. It is worth noting that the value of Fermi length is $\lambda_F\simeq 37$\,nm at $V_{BG}=-1$\,V, this means that the number of channels in the nanowire is $N_{channel}\simeq\pi d_{wire}^2/\lambda_F^2\simeq 24$, and the localization length is $l_{loc}\simeq N_{channel}l_e\simeq 240$\,nm. This value is comparable to $l_{seg}$ and to the actual phase-coherence length for $V_{BG}=-1$\,V, i.e. $l_\phi(V_{BG} =-1$V$) = [(l_e(V_{BG} = -1$V$)/l_e(V_{BG}=10$V$)]^{0.5}\cdot200$\,nm$=160$\,nm. Thus, the many body localization is possible in each segment ($l_\phi(V_{BG} =-1$\,V$)  \simeq l_{loc}$). Finally, when $k_BT$ is the largest energy scale ($k_BT>E_{Th,wire} \sim 3\mu$eV), a simple exponential decay of the quasiparticle life time is set again, here $E_{Th,wire}$ is the Thouless energy of the whole wire. This regime corresponds to the homogenous diffusive transport regime in the current work. 
	
	Thus, as in AGKL theoretical picture so in the current experiment there is a special regime with non-trivial behavior of the electronic system wave function or fractal  Bownian motion of magnetoconductance curve lying in between Coulomb blockade and homogenous diffusion transport regimes.  
		
	There are quite a number of theoretical papers in which quasi-localized states formed by resonant scatters are considered as a possible origin of formation of the observed fractional Bownian motion of magnetoconductance dependence in the {\it intermediate regime} \cite{Altshuler1987, Muzykantskii1995, Mirlin1995}. 
	However, there have been no clear experimental evidences or visualizations of such kind of states. Here we show as the SGM mapping scans visualize them and hence the fractional Bownian motion they initiate.
	
	\section{Conclusion}
    We performed the low temperature measurements of the magnetotransport in doped InAs nanowire in the presence of a charged tip of an atomic force microscope serving as a mobile gate. By varying the carrier concentration, the system under investigation passes through four transport regimes from Coulomb blockade to homogeneous diffusive transport. Fractional Bownian motion of magnetoconductance dependence is observed in non-linear and linear resonant transport regimes. Two key experiments for demonstration of transition from the linear resonant transport regime to the diffusive one are presented. The experimental results are in general agreement with the theoretical model proposed by AGKL.
	\section{Acknowledgements}
	
	The authors would like to thank Chrisian Bl\"omers for samples preparation and Raffaella Calarco for nanowire growth. This work is supported by the RSF grant 23-22-00141, https://rscf.ru/project/23-22-00141/. \\
	
	{}
	
	\begin{figure}
		\includegraphics*[width=0.4\columnwidth]{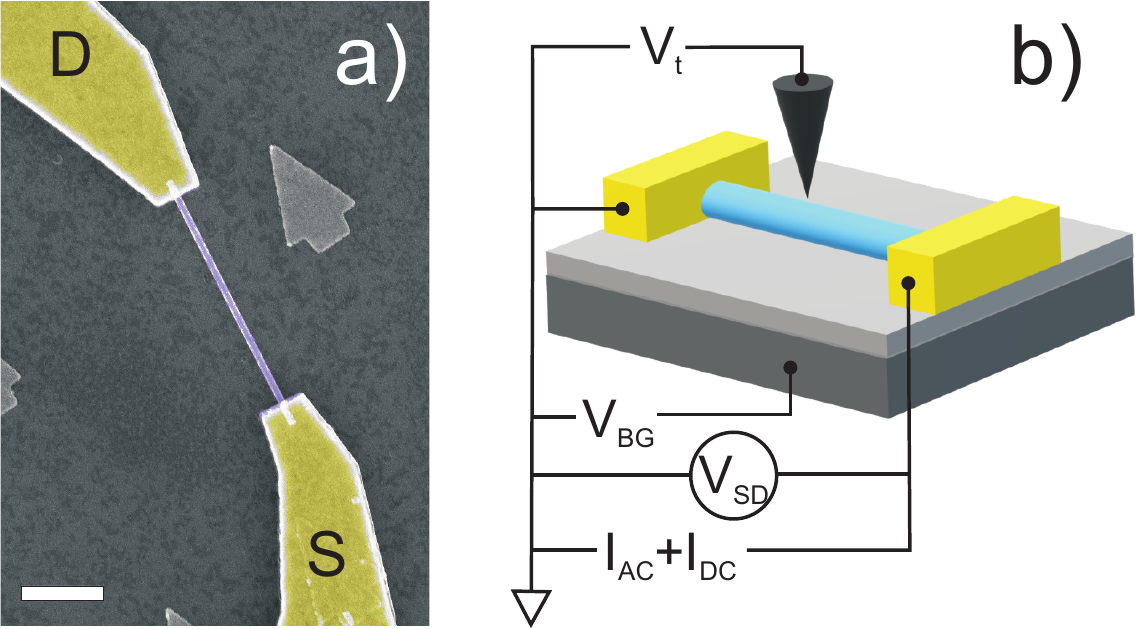}
		\caption {(color online) a) False color scanning electron microscope image of the InAs nanowire. The scale bar corresponds to 1\,$\mu$m. b) Schematic diagram of the experimental setup. In the Figures, the nanowire has a blue color and two metallic contact pads are yellow, source and drain contacts are labelled S and D, respectively.} \label{Fig1}
	\end{figure}
	
	\begin{figure}
		\includegraphics*[width=0.4\columnwidth]{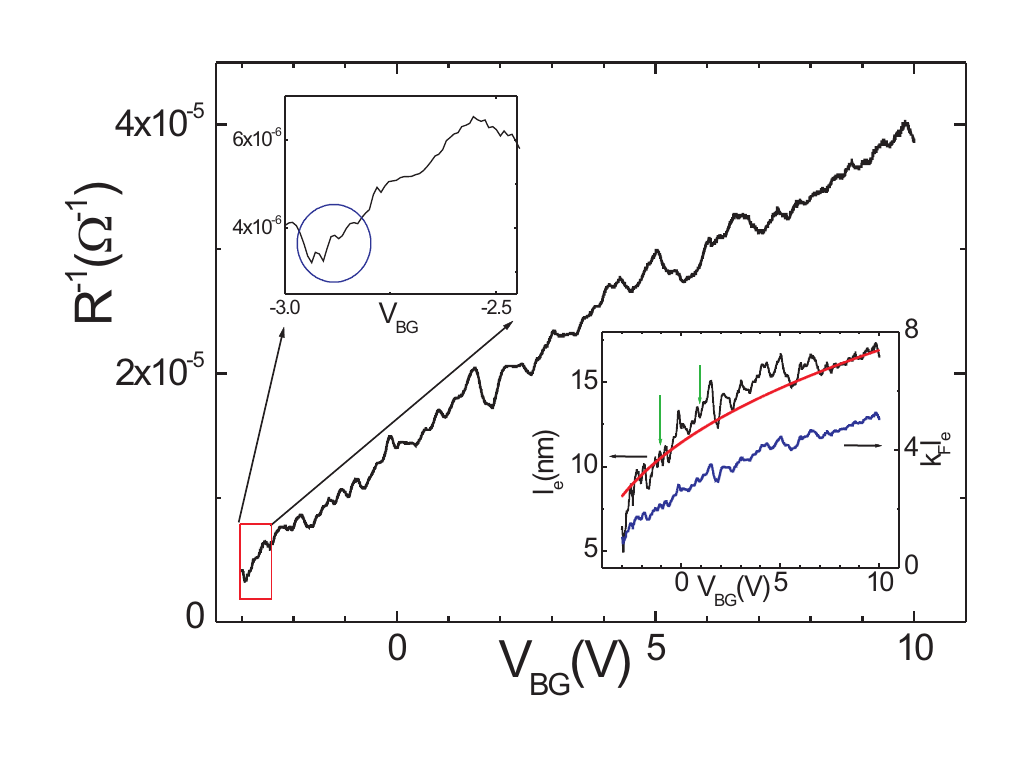}
		\caption {(color online) Conductance $R^{-1}$ of the InAs nanowire versus back gate voltage ($V_{BG}$). Non-regular oscillations come from universal conductance fluctuations. The upper-left insert shows the dependence of the conductance on the back gate voltage in the range indicated by the dashed frame in the Figure. Small period oscillations within the range marked with blue circle come from residual Coulomb blockade of the quantum dot positioned in the mid part of the nanowire (see main text for details). The lower-right insert presents the elastic scattering length as a function of the back gate voltage (black curve, left scale). Blue curve in the insert shows the dependence of $k_Fl_e$ on the back gate voltage (right scale). Two green arrows point to $V_{BG}=-1$\,V and $V_{BG}=1$\,V.}
		\label{Fig2}
	\end{figure}
	
	\begin{figure}
		\includegraphics*[width=0.4\columnwidth]{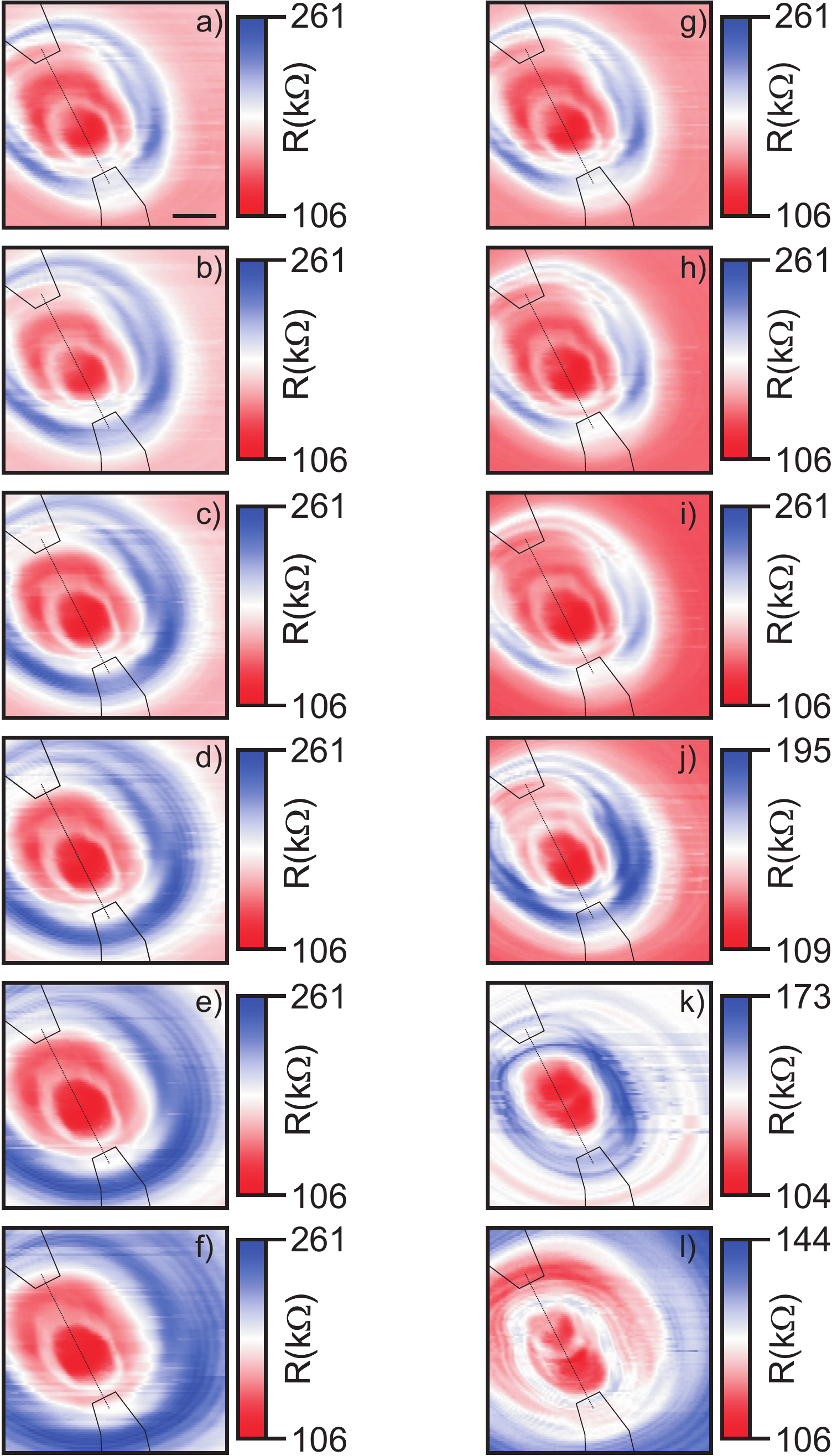}
		\caption {(color online) a)-f) Scanning gate microscopy maps of InAs nanowire carried out at zero magnetic field and in range of back gate voltages varied from $V_{BG}=-3.00$\,V to $-2.95$\,V with step of 10\,mV, respectively. g)-l) Scanning gate microscopy maps gained at $V_{BG}=-3.0$\,V and $B=0.1$\,T, 0.2\,T, 0.3\,T, 0.5\,T, 1.0\,T, and 2.0\,T, respectively. Concentric circles are the sign of the Coulomb blockade in quantum dot positioned at the mid part of the nanowire. The scale bar in Figure a) corresponds to 1\,$\mu$m. The scale and the area of scanning gate microscopy maps are the same for all scans. The black solid lines mark the borders of contact pads and the dotted line marks the axis of nanowire. }
		\label{Fig3}
	\end{figure}
	
	\begin{figure}
		\includegraphics*[width=0.4\columnwidth]{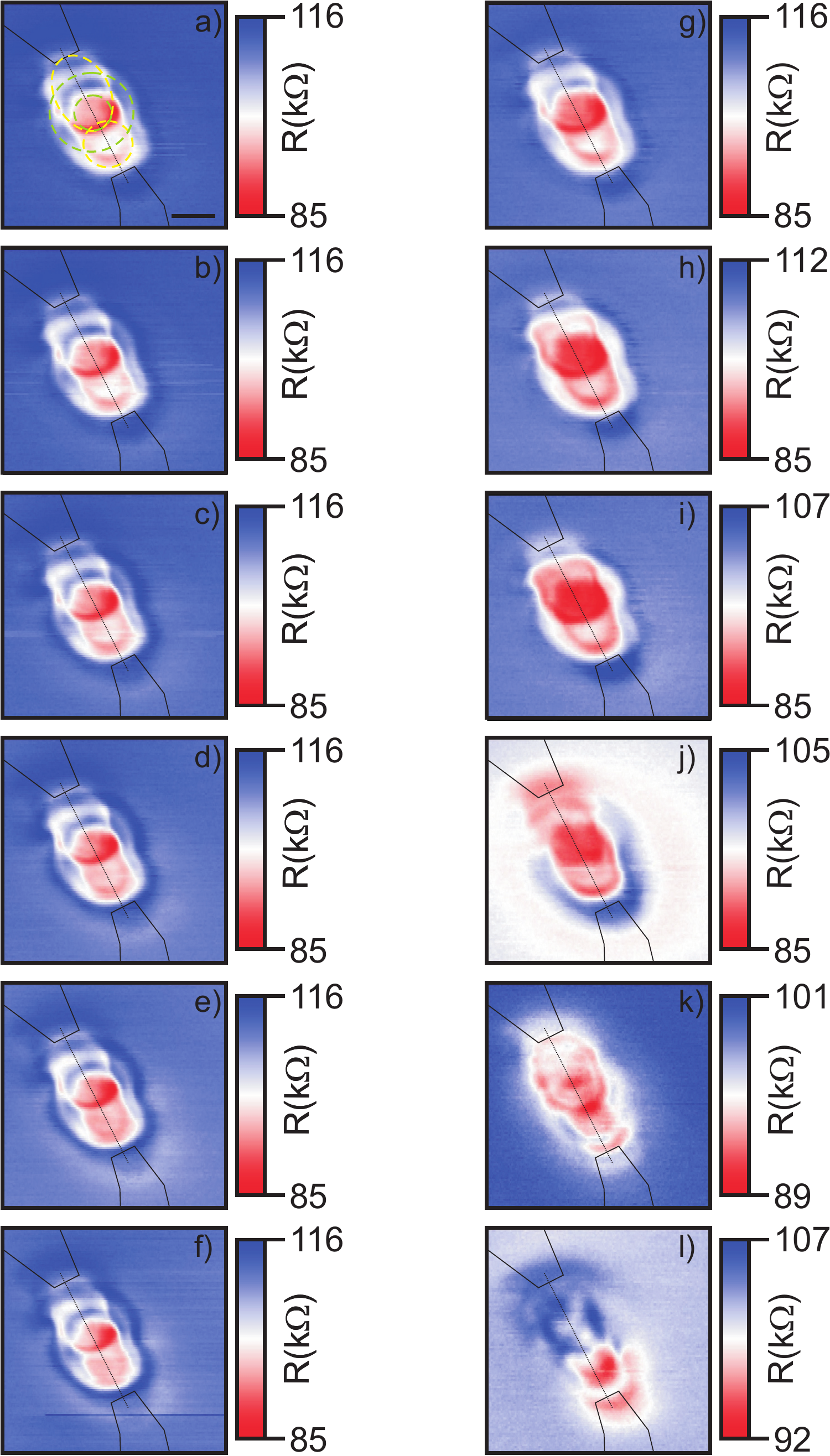}
		\caption {(color online) a)-f) Scanning gate microscopy maps of InAs nanowire obtained at zero magnetic field and in range of back gate voltages varied from $V_{BG}=-2.10$\,V to $-2.15$\,V with step of 10\,mV, respectively. g)-l) Scanning gate microscopy maps gained at $V_{BG}=-2.10$\,V and $B=0.1$\,T, 0.2\,T, 0.3\,T, 0.5\,T, 1.0\,T, and 2.0\,T, respectively. The scale bar in Figure~4a) corresponds to 1\,$\mu$m. The scale and the area of scanning gate microscopy maps are the same for all scans. The black solid lines mark the borders of contact pads and the dotted line marks the axis of nanowire. Ovals and circles of equicapacitance for all sections of the nanowire are shown in the Figure ~a). Two yellow ovals relate to the Sections I and III, two green circles mark equicapacitance curve for central quantum dot (Section II).}
		\label{Fig4}
	\end{figure}
	
	\begin{figure}
		\includegraphics*[width=0.4\columnwidth]{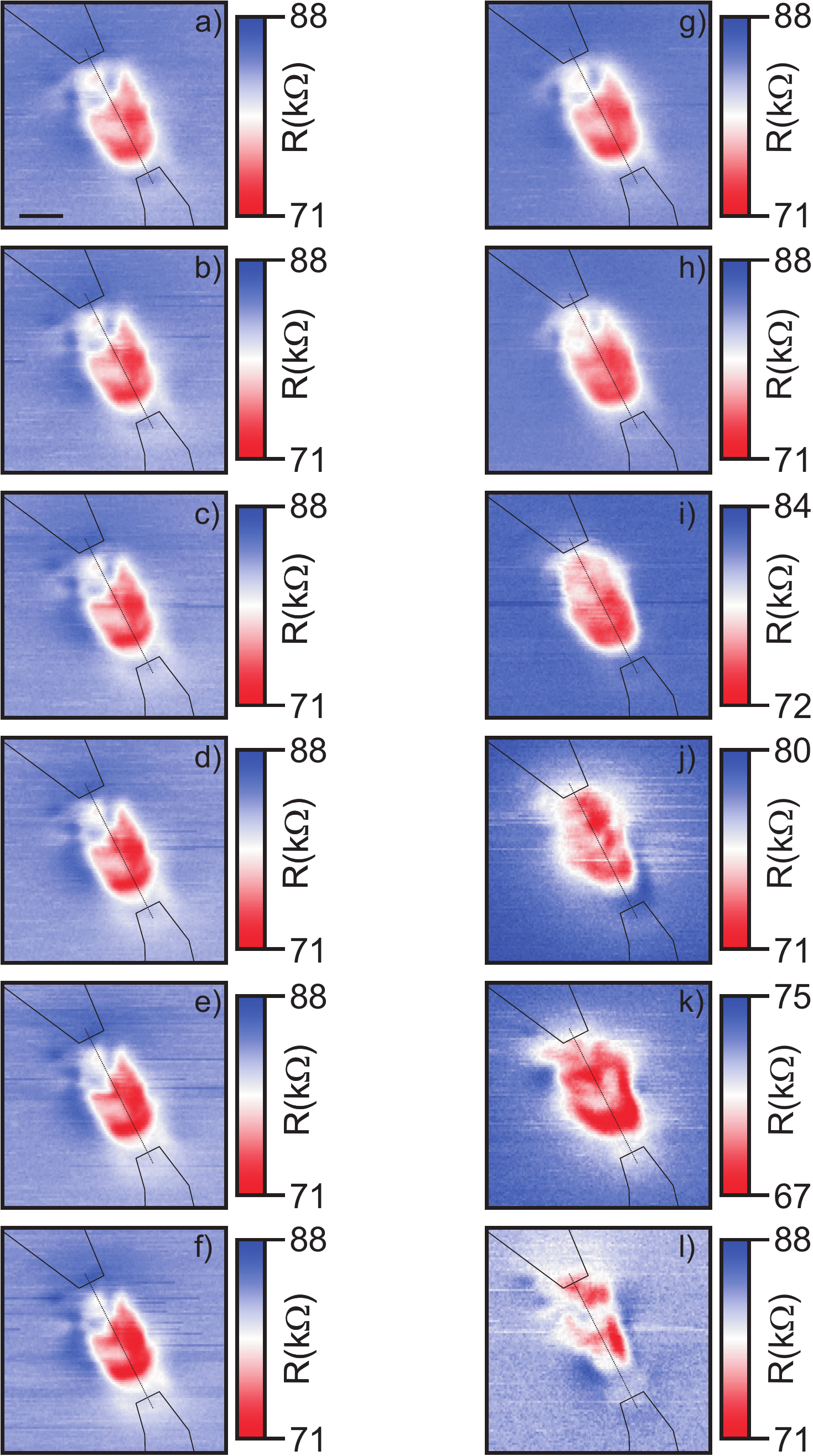}
		\caption {(color online) a)-f) Scanning gate microscopy maps of InAs nanowire made at zero magnetic field and in range of back gate voltages varied from $V_{BG}=-1.00$\,V to $-1.05$\,V with step of 10\,mV, respectively. g)-l) Scanning gate microscopy maps carried out at $V_{BG}=-1.00$\,V and $B=0.1$\,T, 0.2\,T, 0.3\,T, 0.5\,T, 1.0\,T, and 2.0\,T, respectively. The scale bar in Figure a) corresponds to 1\,$\mu$m. The scale and the area of scanning gate microscopy maps are the same for all scans. The black solid lines mark the borders of contact pads and the dotted line marks the axis of nanowire.}
		\label{Fig5}
	\end{figure}
	
	\begin{figure}
		\includegraphics*[width=0.4\columnwidth]{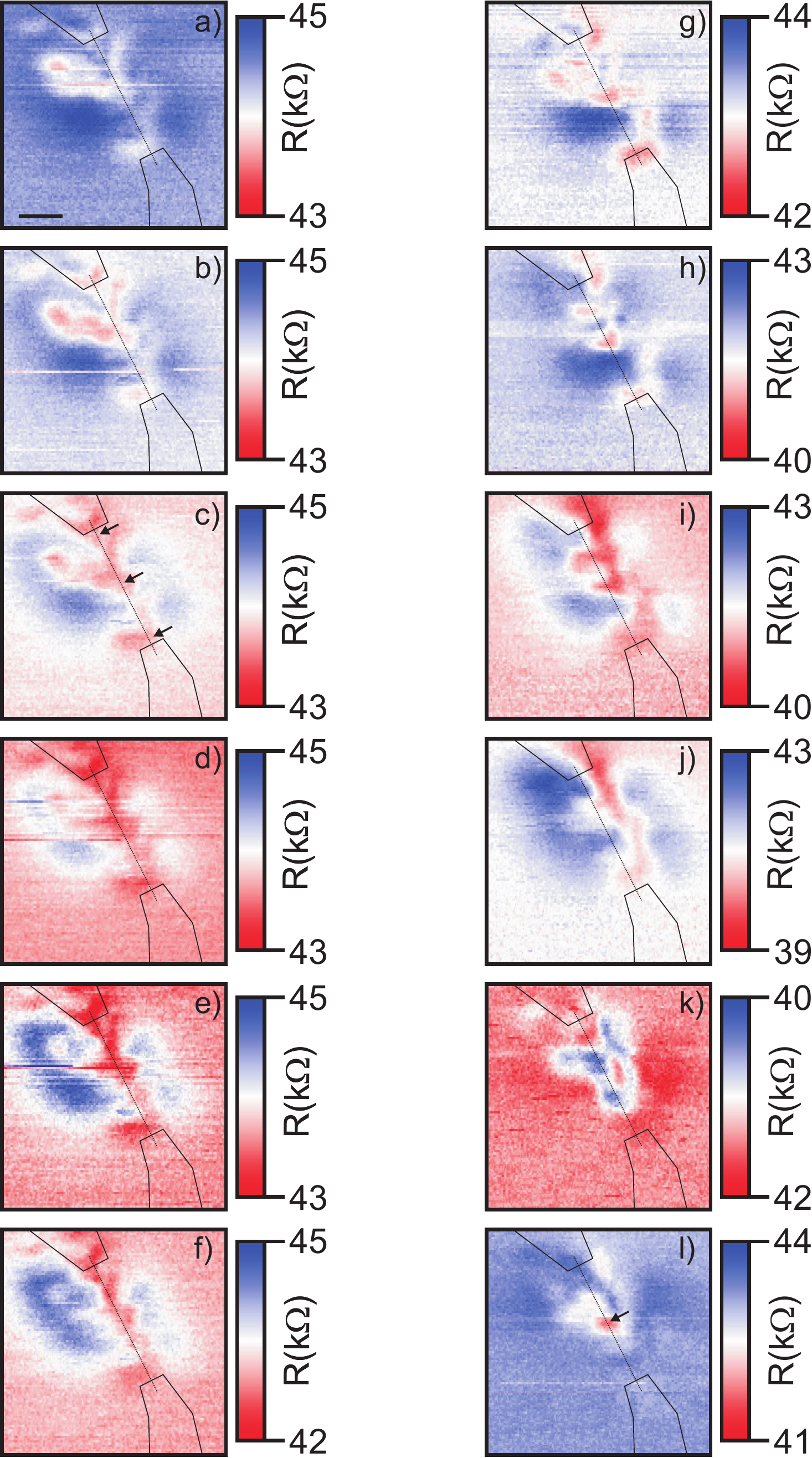}
		\caption {(color online) a)-f) Scanning gate microscopy maps of InAs nanowire obtained at zero magnetic field and in range of back gate voltages varied from $V_{BG}=1.00$\,V to $1.05$\,V with step of 10\,mV, respectively. g)-l) Scanning gate microscopy maps gained at $V_{BG}=1.00$\,V and $B=0.1$\,T, 0.2\,T, 0.3\,T, 0.5\,T, 1.0\,T, and 2.0\,T, respectively. The scale bar in Fig. 6a) corresponds to 1\,$\mu$m. The scale and the area of scanning gate microscopy maps are the same for all scans. The black solid lines mark the borders of contact pads and the dotted line marks the axis of nanowire. Arrows in Figure c) point to the positions of all potential barriers. The arrow in Figure~l) points to the position of the central double barrier.
		}
	\end{figure}
	
	\begin{figure}
		\includegraphics*[width=0.4\columnwidth]{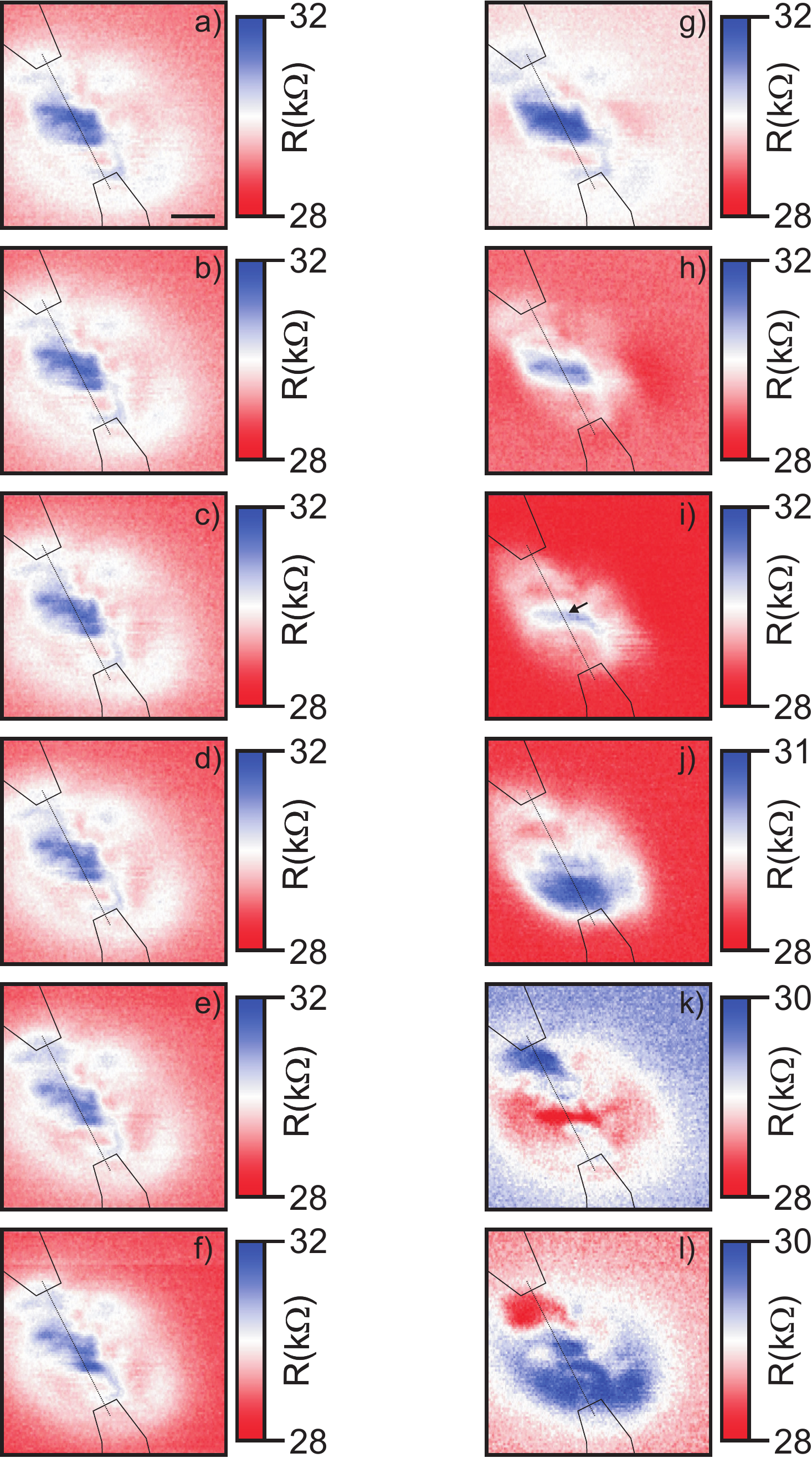}
		\caption {(color online) a)-f) Scanning gate microscopy maps of InAs nanowire carried out at zero magnetic field and in range of back gate voltages varied from $V_{BG}=4.00$\,V to $4.05$\,V with step of 10\,mV, respectively g)-l) Scanning gate microscopy maps performed at $V_{BG}=4.00$\,V and $B=0.1$\,T, 0.2\,T, 0.3\,T, 0.5\,T, 1.0\,T, and 2.0\,T, respectively. The scale bar in Figure a) corresponds to 1\,$\mu$m. The scale and the area of scanning gate microscopy maps are the same for all scans. Tbe black solid lines mark the borders of contact pads and the dotted line marks the axis of nanowire. The arrow in Fig.~i) points to the position of the central double barrier.}
	\end{figure}
	
	\begin{figure}
		\includegraphics*[width=0.4\columnwidth]{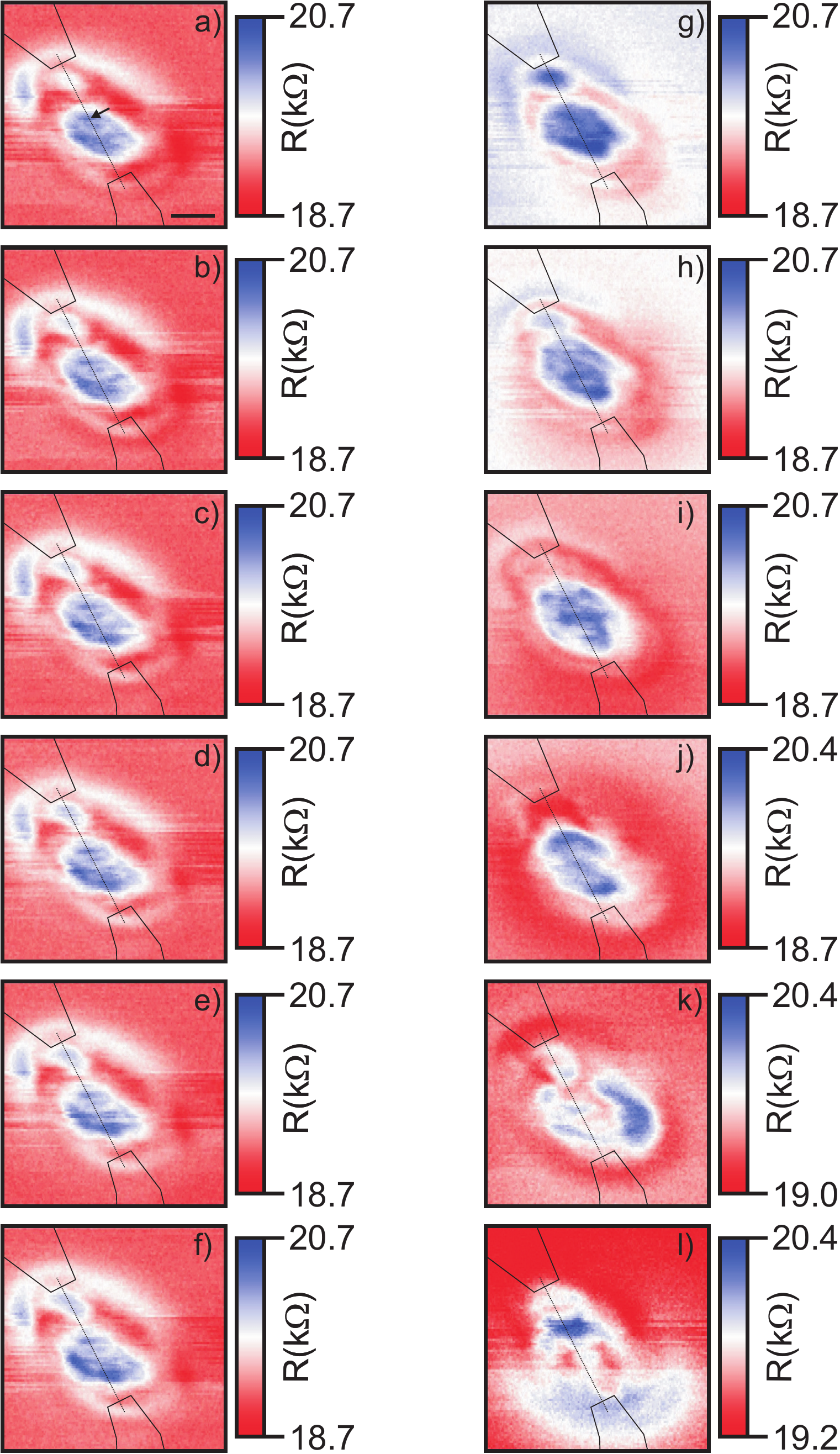}
		\caption {(color online) a)-f) Scanning gate microscopy maps of InAs nanowire obtained at zero magnetic field and in range of back gate voltages varied from $V_{BG}=10.00$\,V to $9.95$\,V with step of 10\,mV, respectively. g)-l) Scanning gate microscopy maps carried out at $V_{BG}=10.00$\,V and $B=0.1$\,T, 0.2\,T, 0.3\,T, 0.5\,T, 1.0\,T, and 2.0\,T, respectively. The scale bar in Figure a) corresponds to 1\,$\mu$m. The scale and the area of scanning gate microscopy maps are the same for all scans. The black solid lines mark the borders of contact pads and the dotted line marks the axis of nanowire. The arrow in Figure ~a) marks the position of the double barrier.
		}
	\end{figure}
	
	\begin{figure}
		\includegraphics*[width=0.4\columnwidth]{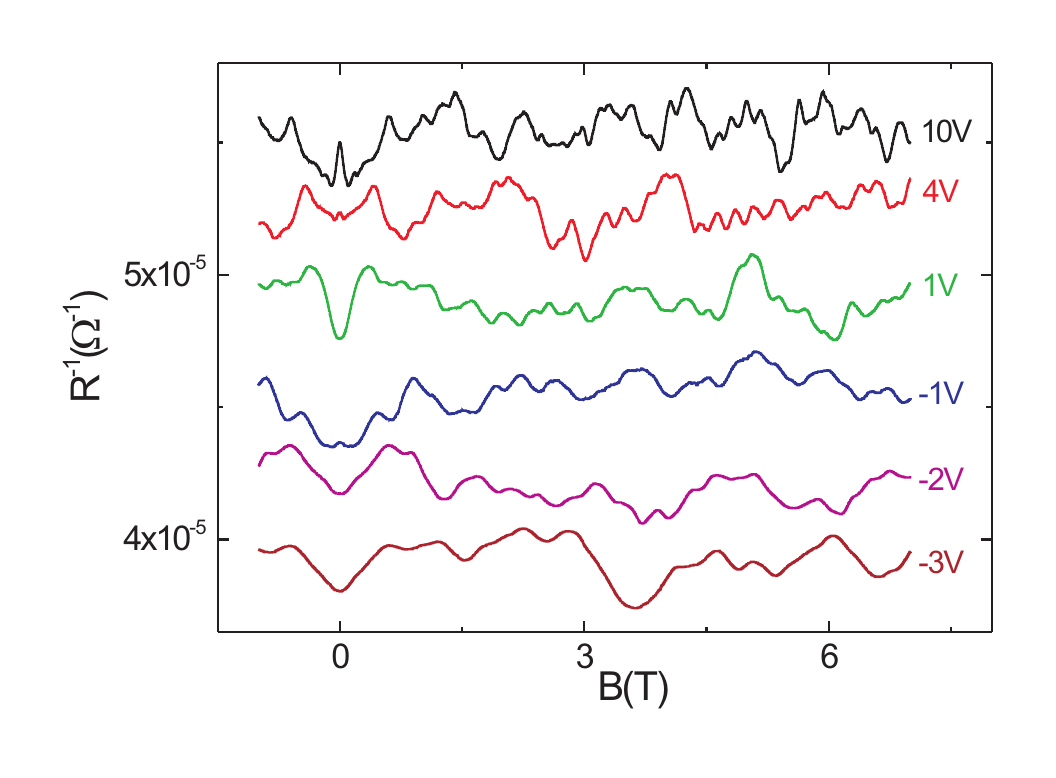}
		\caption {(color online) Magnetic field dependencies of the conductance of the nanowire ($R^{-1}$) measured in the range of magnetic fields from -1\,T to 7\,T and at different back gate voltages as indicated in Figure.
		}
	\end{figure}
	
	\begin{figure}
		\includegraphics*[width=0.4\columnwidth]{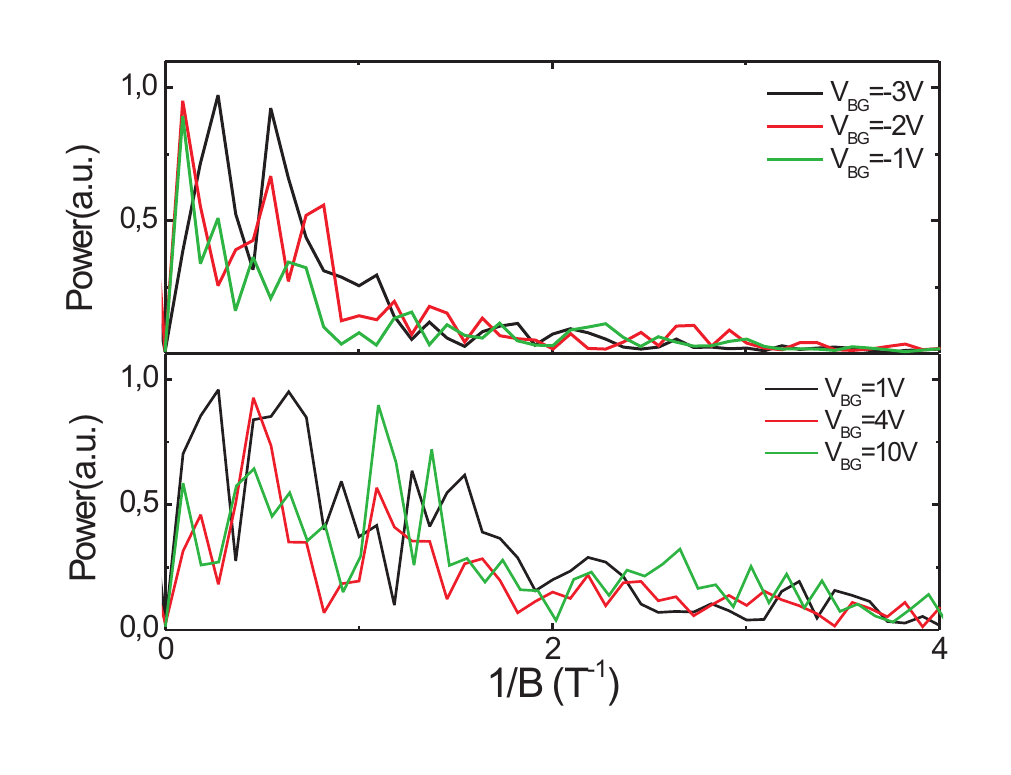}
		\caption {(color online) Top panel is power spectra of magnetoconductance for $V_{BG}=-3$\,V, -2\,V, and -1\,V, see Fig. 9. Bottom panel is power spectra of magnetoconductance for $V_{BG}=1$\,V, 4\,V, and 10\,V, see Fig. 9. The significant suppression of all spectra for $1/B>1$\,T$^{-1}$ presented in the top panel is clearly visible.
		}
%	I would suggest to mark the top panel as a) and the bottom panel as b) 
%	In this case, the Figure captions could be 
%	" a) Power spectra of magnetoconductance for $V_{BG}=-3$\,V, -2\,V and -1\,V, see Fig. 9, b) Power spectra of magnetoconductance for $V_{BG}=1$\,V, 4\,V and 10\,V, see Fig. 9. The significant suppression of all spectra for $1/B>1$\,T$^{-1}$ presented in the figure  a) is clearly visible. 
	
	\end{figure}
	
	\begin{figure}
		\includegraphics*[width=0.4\columnwidth]{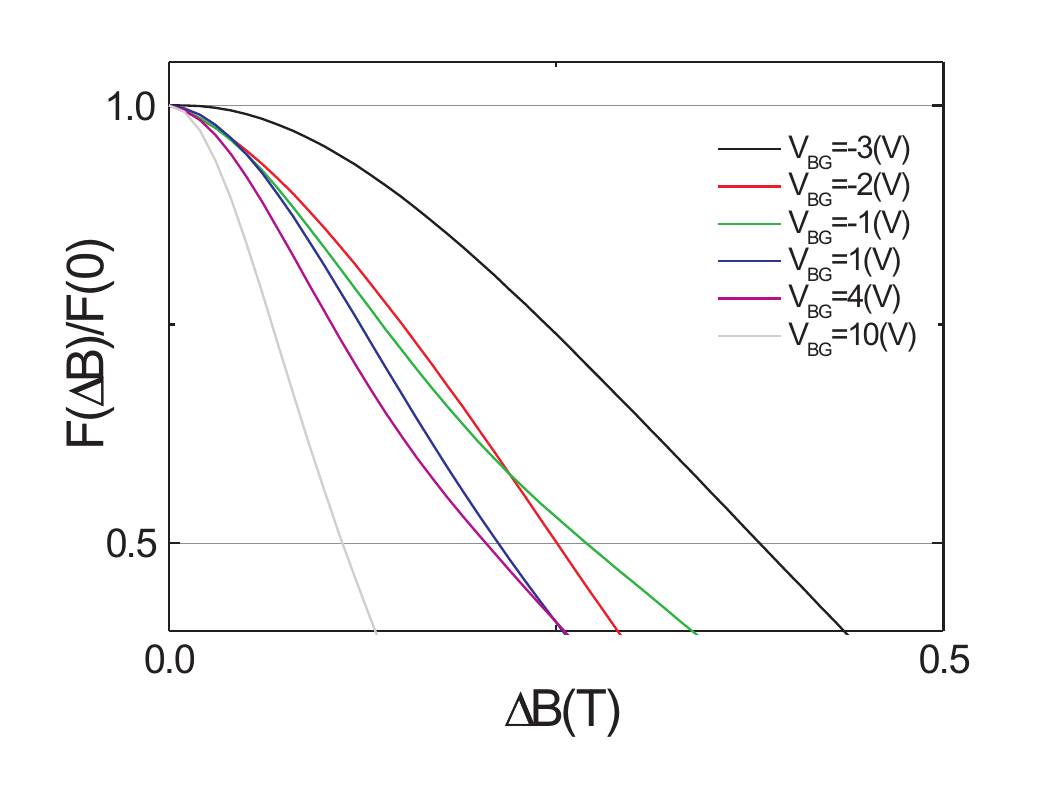}
		\caption {(color online) Normalized correlation functions $F(\Delta B)/F(0)$  calculated from magnetoconductance dependencies presented in Fig. 9 for six different back gate voltages.
		}
	\end{figure}
	
	\begin{figure}
		\includegraphics*[width=0.4\columnwidth]{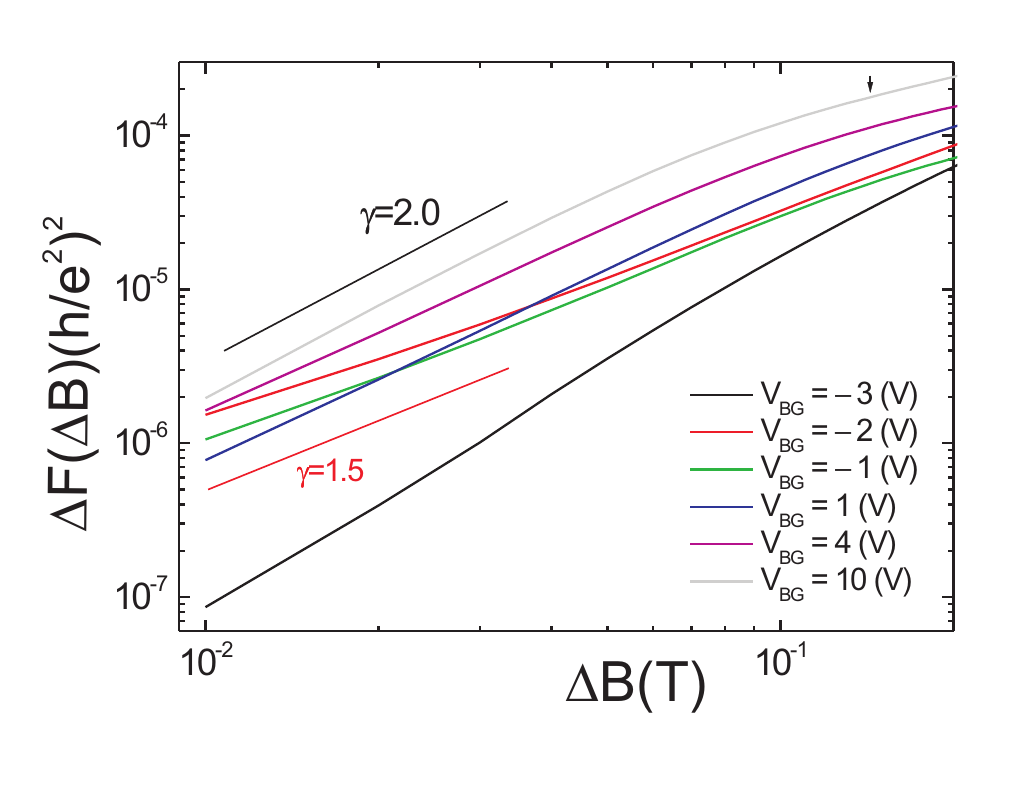}
		\caption {(color online) Deviation of the correlation function for six different back gate voltages. The black and the red straight lines show slopes related to $\gamma = 2.0$ and 1.5, respectively. The arrow points at $B_c$ for $V_{BG}=10$\,V .
		}
	\end{figure}
	
	\begin{figure}
		\includegraphics*[width=0.4\columnwidth]{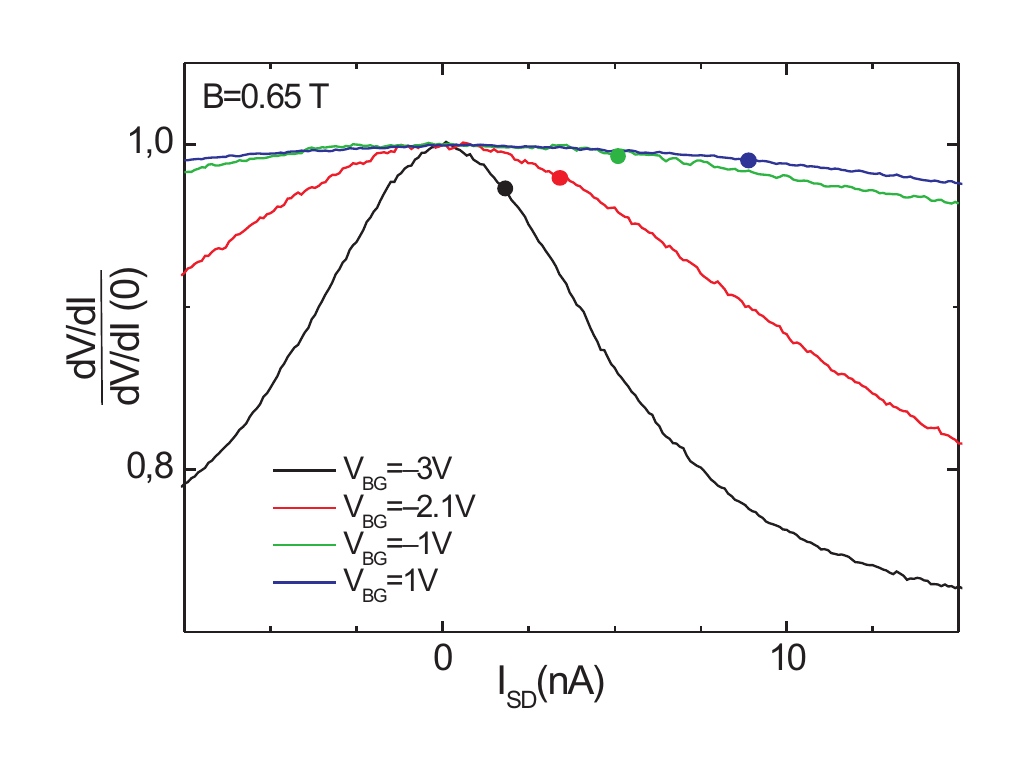}
		\caption {(color online) a) Normalized differential resistance ((dV/dI)/(dV/dI(0))) as a function of the driving current ($I_{SD}$) measured at $V_{BG}=-3$\,V, -2\,V, -1\,V, and 1\,V. The dot in each curve points to the value of the bias current where $eU_{SD}=k_BT$. Measurements were done at $B=0.65$\,T.
		}
	\end{figure}
	
\end{document}